\documentclass[12pt]{article}
\usepackage{amsmath,amsthm}
\usepackage{eucal}
\usepackage{eufrak}
\usepackage{pb-diagram,lamsarrow,pb-lams}

\DeclareMathAlphabet{\bb}{U}{msb}{m}{n}
\gdef\C{\bb C}
\gdef\dZ{\bb Z}

\gdef\R{\bb R}

\DeclareMathOperator{\spin}{{\bf Spin}}

\newcommand{\re}{\mbox{\rm Re}\,}

\newcommand{\cL}{\mathcal{L}}
\newcommand{\cP}{\mathcal{P}}

\newcommand{\sz}{{\sf z}}
\newcommand{\sL}{\Lambda}

\newcommand{\bp}{{\bf p}}

\newcommand{\fM}{\mathfrak{M}}
\newcommand{\fF}{\mathfrak{F}}

\newcommand{\fL}{\mathfrak{L}}

\newcommand{\fg}{\mathfrak{g}}

\newcommand{\balpha}{\boldsymbol{\alpha}}
\newcommand{\cl}{C\kern -0.2em \ell}

\newcommand{\hypergeom}[5]{\mbox{$
_#1 F_#2\left.
\!\!
\left(
\!\!\!\!
\begin{array}{c}
\multicolumn{1}{c}{\begin{array}{c}
#3
\end{array}}\\[1mm]
\multicolumn{1}{c}{\begin{array}{c}
#4
\end{array}}\end{array}
\!\!\!\!
\right|\displaystyle{#5}\right)
$}
}

\newcommand{\tg}{\tan}
\newcommand{\ch}{\cosh}

\newcommand{\tnh}{\tanh}

\begin{document}
\title{Relativistic wavefunctions on the Poincar\'{e} group}
\author{V. V. Varlamov\\
{\small Department of Mathematics, Siberia State University of Industry,}\\
{\small Kirova 42, Novokuznetsk 654007, Russia}}
\date{}
\maketitle
\begin{abstract}
The Biedenharn type relativistic wavefunctions are considered on the group
manifold of the Poincar\'{e} group. It is shown that the wavefunctions
can be factorized on the group manifold into translation group and
Lorentz group parts. A Lagrangian formalism and field equations for such
factorizations are given. Parametrizations of the functions obtained are
studied in terms of a ten-parameter set of the Poincar\'{e} group.
An explicit construction of the wavefunction for the spin 1/2 is given.
A relation of the proposed description with the quantum field theory and
harmonic analysis on the Poincar\'{e} group is discussed.
\end{abstract}
PACS {\bf numbers}: 03.65.Pm, 03.65.Ge, 02.30.Gp
\section{Introduction}
In 1988, Biedenharn {\em et al.} \cite{BBTD88} introduced the Poincar\'{e} group
representations of wavefunctions on the space of complex spinors.
The construction presented in \cite{BBTD88} is an extension of the
Wigner's group theoretical method \cite{Wig39}. On the other hand, in
accordance with basic principles of quantum field theory, a wavefunction of
the particle is a solution of some relativistic wave equation. Relativistic
wavefunctions of the work \cite{BBTD88} were introduced without explicit reference
to wave equations and for this reason they represent purely group
theoretical constructions. However, if we further develop the Poincar\'{e}
group representations of wavefunctions with reference to such basic notions
of QFT as a Lagrangian and wave equations, then we come to a quantum field
theory on the Poincar\'{e} group\footnote{In 1955, Finkelstein \cite{Fin55}
showed that elementary particle models with internal  degrees of freedom
can be described on manifolds larger then Minkowski spacetime (homogeneous
spaces of the Poincar\'{e} group). A consideration of the field models
on the homogeneous spaces leads to a generalization of the concept of
wavefunction. One of the first examples of such generalized wavefunctions
was studied by Nilsson and Beskow \cite{NB67}.}
(QFTPG) introduced by Lur\c{c}at
in 1964 \cite{Lur64} (see also \cite{BK69,Kih70,BF74,Aro76,KLS95,Tol96,LSS96,Dre97,GS01,GL01}
and references therein). In contrast to the standard QFT (QFT in the
Minkowski spacetime) case, for QFTPG all the notions and quantities are
constructed on a ten-dimensional group manifold $\fF$ of the Poincar\'{e}
group. It should
be noted here that a construction of relativistic wave equations theory
on the group manifold $\fF$ is one of the primary problems in this
area, which remains incompletely solved. Wavefunctions and
wave equations on a six-dimensional submanifold $\fL\subset\fF$, which is
a group manifold of the Lorentz group, have been studied in the recent
work \cite{Var031}.

In the present paper we consider Biedenharn type relativistic wavefunctions
on the group manifold $\fF$. It is shown that the general form of the
wavefunctions inherits its structure from the semidirect product
$SL(2,\C)\odot T_4$ and for that reason the wavefunctions on $\fF$ are
represented by a factorization $\psi(x)\psi(\fg)$, where $x\in T_4$,
$\fg\in SL(2,\C)$. Using a Lagrangian formalism on the tangent bundle
$T\fF$ of the manifold $\fF$, we obtain field equations separately for the
parts $\psi(x)$ and $\psi(\fg)$. Solutions of the field equations for
$\psi(x)$ can be obtained via the usual plane-wave approximation.
In turn, solutions of the field equations with $\psi(\fg)$ have been found
in the form of expansions in associated 
hyperspherical functions\footnote{Matrix elements of both spinor and
principal series representations of the Lorentz group are expressed via
the hyperspherical functions \cite{Var022,Var042}.}.
The wavefunction on the Poincar\'{e} group in the case of
$(1/2,0)\oplus(0,1/2)$ representation space, usually related with
electron-positron field, is considered by way of example.

\section{Preliminaries}
Let us consider some basic facts concerning the Poincar\'{e} group $\cP$.
First of all, the group $\cP$ has the same number of connected components
as the Lorentz group. Later on we will consider only the component
$\cP^\uparrow_+$ corresponding the connected component
$L^\uparrow_+$ (the so-called special Lorentz group \cite{RF}). As is known,
a universal covering $\overline{\cP^\uparrow_+}$ of the group $\cP^\uparrow_+$
is defined by a semidirect product
$\overline{\cP^\uparrow_+}=SL(2,\C)\odot T_4\simeq\spin_+(1,3)\odot T_4$, 
where $T_4$ is a subgroup of four-dimensional translations. The relations
between the groups
$\overline{\cP^\uparrow_+}$, $\cP^\uparrow_+$, $SL(2,\C)$ and
$L^\uparrow_+$ are defined by the following diagram of exact sequences:
\[
\dgARROWLENGTH=1em
\begin{diagram}
\node[3]{1}\arrow{s}\node{1}\arrow{s}\\
\node[3]{\dZ_2}\arrow{s}\node{\dZ_2}\arrow{s}\\
\node{1}\arrow{e}\node{T_4}\arrow{e}\node{\overline{\cP^\uparrow_+}}\arrow{s}
\arrow{e,<>}\node{SL(2,\C)}\arrow{s}\arrow{e,<>}\node{1}\\
\node{1}\arrow{e}\node{T_4}\arrow{e}\node{\cP^\uparrow_+}\arrow{s}
\arrow{e,<>}\node{L^\uparrow_+}\arrow{s}\arrow{e,<>}\node{1}\\
\node[3]{1}\node{1}
\end{diagram}
\]
The diagram shows that $\overline{\cP^\uparrow_+}$
($\cP^\uparrow_+$) is the semidirect product of $SL(2,\C)$
($L^\uparrow_+$) and $T_4$.

The each transformation $T_{\balpha}\in\cP^\uparrow_+$ is defined by
a parameter set $\balpha(\alpha_1,\ldots,\alpha_{10})$, which can be 
represented by a point of the space $\fF_{10}$. The space $\fF_{10}$
possesses locally euclidean properties; therefore, it is a manifold, called
{\it a group manifold of the Poincare group}. It is easy to see that the
set $\balpha$ can be divided into two subsets,
$\balpha(x_1,x_2,x_3,x_4,\fg_1,\fg_2,\fg_3,\fg_4,\fg_5,\fg_6)$, where
$x_i\in T_4$ are parameters of the translation subgroup,
$\fg_j$ are parameters of the group $SL(2,\C)$. In turn, the transformation
$T_{\fg}$ is defined by a set $\fg(\fg_1,\ldots,\fg_6)$, which can be
represented by a point of a six-dimensional submanifold $\fL_6\subset\fF_{10}$,
called {\it a group manifold of the Lorentz group}.

In the present paper we restrict ourselves to consideration of finite
dimensional representations of the Poincar\'{e} group. The group $T_4$ of
four-dimensional translations is an Abelian group, formed by a direct
product of the four one-dimensional translation groups, each of which
is isomorphic to an additive group of real numbers. Hence it follows that all
irreducible representations of $T_4$ are one dimensional and expressed
via the exponential. In turn, as shown by Naimark \cite{Nai58},
spinor representations exhaust all the finite dimensional irreducible
representations of the group $SL(2,\C)$. Any spinor representation of
$SL(2,\C)$ can be defined in the space of symmetric polynomials in the
following form:
\begin{gather}
p(z_0,z_1,\bar{z}_0,\bar{z}_1)=\sum_{\substack{(\alpha_1,\ldots,\alpha_k)\\
(\dot{\alpha}_1,\ldots,\dot{\alpha}_r)}}\frac{1}{k!\,r!}
a^{\alpha_1\cdots\alpha_k\dot{\alpha}_1\cdots\dot{\alpha}_r}
z_{\alpha_1}\cdots z_{\alpha_k}\bar{z}_{\dot{\alpha}_1}\cdots
\bar{z}_{\dot{\alpha}_r}\label{SF}\\
(\alpha_i,\dot{\alpha}_i=0,1),\nonumber
\end{gather}
%\begin{sloppypar}\noindent
where the numbers 
$a^{\alpha_1\cdots\alpha_k\dot{\alpha}_1\cdots\dot{\alpha}_r}$
are unaffected at the permutations of indices. 
The expressions (\ref{SF}) can be understood as {\it functions on the
Lorentz group}.
When the coefficients
$a^{\alpha_1\cdots\alpha_k\dot{\alpha}_1\cdots\dot{\alpha}_r}$ in
(\ref{SF}) depend on the variables $x_i\in T_4$ ($i=1,2,3,4$),
we come to the Biedenharn type functions \cite{BBTD88}:
\begin{gather}
p(x,z,\bar{z})=\sum_{\substack{(\alpha_1,\ldots,\alpha_k)\\
(\dot{\alpha}_1,\ldots,\dot{\alpha}_r)}}\frac{1}{k!\,r!}
a^{\alpha_1\cdots\alpha_k\dot{\alpha}_1\cdots\dot{\alpha}_r}(x)
z_{\alpha_1}\cdots z_{\alpha_k}\bar{z}_{\dot{\alpha}_1}\cdots
\bar{z}_{\dot{\alpha}_r}.\label{SF2}\\
(\alpha_i,\dot{\alpha}_i=0,1)\nonumber
\end{gather}
The functions (\ref{SF2}) should be considered as {\it the functions on
the Poincar\'{e} group}\footnote{Antisymmetrizing the variables
$z$ and $\bar{z}$ in (\ref{SF2}), we come to the functions on the supergroup.
In particular case of Grassmann variables \cite{Ber87} we have the functions
on a so-called super-Poincar\'{e} group.}.
Some applications of these functions are contained
in \cite{Vas96,GS01}. Representations of 
the Poincar\'{e} group
$SL(2,\C)\odot T(4)$ are realized via the functions (\ref{SF2}).

\section{Field equations on the Poincar\'{e} group}
Let $\cL(\balpha)$ be a Lagrangian on the group manifold $\fF$ of the
Poincar\'{e} group (in other words, $\cL(\balpha)$ is a ten-dimensional
point function), where $\balpha$ is the parameter set of this group.
Then we will call
an integral for $\cL(\balpha)$ on some 10-dimensional volume $\Omega$
of the group manifold {\it an action on the Poincar\'{e}
group}:
\[
A=\int\limits_\Omega d\balpha\cL(\balpha),
\]
where $d\balpha$ is a Haar measure on the group $\cP$.

Let $\psi(\balpha)$ be a function on the group manifold $\fF$ (now it is
sufficient to assume that $\psi(\balpha)$ is a square integrable function
on the Poincar\'{e} group) and let
\begin{equation}\label{ELE}
\frac{\partial\cL}{\partial\psi}-\frac{\partial}{\partial\balpha}
\frac{\partial\cL}{\partial\frac{\partial\psi}{\partial\balpha}}=0
\end{equation}
be Euler-Lagrange equations on $\fF$ (more precisely speaking, the equations
(\ref{ELE}) act on the tangent bundle 
$T\fF=\underset{\balpha\in\fF}{\cup}T_{\balpha}\fF$ of the manifold $\fF$;
see \cite{Arn}). Let us introduce a Lagrangian $\cL(\balpha)$ depending on
the field function $\psi(\balpha)$ as follows:
\[
\cL(\balpha)=-\frac{1}{2}\left(\psi^\ast(\balpha)B_\mu
\frac{\partial\psi(\balpha)}{\partial\balpha_\mu}-
\frac{\partial\psi^\ast(\balpha)}{\partial\balpha_\mu}B_\mu\psi(\balpha)\right)
-\kappa\psi^\ast(\balpha)B_{11}\psi(\balpha),
\]
where $B_\nu$ ($\nu=1,2,\ldots,10$) are square matrices. The number of
rows and columns in these matrices is equal to the number of components
of $\psi(\balpha)$; $\kappa$ is a non-null real constant.

Further, if $B_{11}$ is non-singular, then we can introduce the matrices
\[
\Gamma_\mu=B^{-1}_{11}B_\mu,\quad \mu=1,2,\ldots,10,
\]
and represent the Lagrangian $\cL(\balpha)$ in the form
\begin{equation}\label{Lagrange}
\cL(\balpha)=-\frac{1}{2}\left(\overline{\psi}(\balpha)\Gamma_\mu
\frac{\partial\psi(\balpha)}{\partial\balpha_\mu}-
\frac{\overline{\psi}(\balpha)}{\partial\balpha_\mu}\Gamma_\mu
\psi(\balpha)\right)-\kappa\overline{\psi}(\balpha)\psi(\balpha),
\end{equation}
where
\[
\overline{\psi}(\balpha)=\psi^\ast(\balpha)B_{11}.
\]
As a direct consequence of (\ref{SF2}), the relativistic wavefunction
$\psi(\balpha)$ on the group manifold $\fF$ is represented by the following
factorization:
\begin{equation}\label{WF}
\psi(\balpha)=\psi(x)\psi(\fg)=\psi(x_1,x_2,x_3,x_4)
\psi(\varphi,\epsilon,\theta,\tau,\phi,\varepsilon),
\end{equation}
where $\psi(x_i)$ is a function depending on the parameters of the subgroup
$T_4$, $x_i\in T_4$ ($i=1,\ldots 4$), and $\psi(\fg)$ is a function on the
Lorentz group, where six parameters of this group are defined by the
Euler angles $\varphi$, $\epsilon$, $\theta$, $\tau$, $\phi$, $\varepsilon$
which compose complex angles of the form $\varphi^c=\varphi-i\epsilon$,
$\theta^c=\theta-i\tau$, $\phi^c=\phi-i\varepsilon$.

Varying $\psi(x)$ and $\overline{\psi}(x)$ independently, we obtain from
(\ref{Lagrange}) in accordance with (\ref{ELE}) the following equations:
\begin{equation}\label{FET}
\begin{array}{ccc}
\Gamma_i\dfrac{\partial\psi(x)}{\partial x_i}+\kappa\psi(x)&=&0,\\
\Gamma^T_i\dfrac{\partial\overline{\psi}(x)}{\partial x_i}-
\kappa\overline{\psi}(x)&=&0.
\end{array}\quad(i=1,\ldots,4)
\end{equation}
Analogously, varying $\psi(\fg)$ and $\overline{\psi}(\fg)$ independently,
one gets
\begin{equation}\label{FEL}
\begin{array}{ccc}
\Gamma_k\dfrac{\partial\psi(\fg)}{\partial\fg_k}+\kappa\psi(\fg)&=&0,\\
\Gamma^T_k\dfrac{\overline{\psi}(\fg)}{\partial\fg_k}-
\kappa\overline{\psi}(\fg)&=&0,
\end{array}\quad(k=1,\ldots,6)
\end{equation}
where
\[
\psi(\fg)=\begin{pmatrix}
\psi(\fg)\\
\dot{\psi}(\fg)
\end{pmatrix},\quad
\Gamma_k=\begin{pmatrix}
0 & \Lambda^\ast_k\\
\Lambda_k & 0
\end{pmatrix}.
\]
The doubling of representations, described by a bispinor
$\psi(\fg)=(\psi(\fg),\dot{\psi}(\fg))^T$, is the well known feature of
the Lorentz group representations \cite{GMS,Nai58}. The structure of the
matrices $\Lambda_k$ and $\Lambda^\ast_k$ is studied 
in details in \cite{Var031}. Since a universal covering $SL(2,\C)$ of the
proper orthochronous Lorentz group is a complexification of the group
$SU(2)$ (see, for example, \cite{Vil68,Var022}), it is more
convenient to express the six parameters $\fg_k$ of the Lorentz group via the
three parameters $a_1$, $a_2$, $a_3$ of the group $SU(2)$. It is obvious that
$\fg_1=a_1$, $\fg_2=a_2$, $\fg_3=a_3$, $\fg_4=ia_1$, $\fg_5=ia_2$,
$\fg_6=ia_3$. Then the first equation from (\ref{FEL}) can be written as
\begin{eqnarray}
\sum^3_{j=1}\Lambda^\ast_j\frac{\partial\dot{\psi}}
{\partial\widetilde{a}_j}+i\sum^3_{j=1}\Lambda^\ast_j
\frac{\partial\dot{\psi}}{\partial\widetilde{a}^\ast_j}+
\kappa^c\psi&=&0,\nonumber\\
\sum^3_{j=1}\Lambda_j\frac{\partial\psi}{\partial a_j}-
i\sum^3_{j=1}\Lambda_j\frac{\partial\psi}{\partial a^\ast_j}+
\dot{\kappa}^c\dot{\psi}&=&0,\label{Complex}
\end{eqnarray}
where $a^\ast_1=-i\fg_4$, $a^\ast_2=-i\fg_5$, $a^\ast_3=-i\fg_6$, and
$\widetilde{a}_j$, $\widetilde{a}^\ast_j$ are the parameters corresponding
the dual basis. In essence, the equations (\ref{Complex}) are defined
in a three-dimensional complex space $\C^3$. In turn, the space $\C^3$
is isometric to a six-dimensional bivector space $\R^6$ (a parameter space
of the Lorentz group \cite{Kag26,Pet69}). The bivector space $\R^6$ is
a tangent space of the group manifold $\fL$ of the Lorentz group; that is,
the manifold $\fL$ at each point is equivalent locally to the
space $\R^6$. Thus, for all $\fg\in\fL$ we have $T_{\fg}\fL\simeq\R^6$.
General solutions of the system (\ref{Complex}) have been found in the
work \cite{Var031}
on the tangent bundle
$T\fL=\underset{\fg\in\fL}{\cup}T_{\fg}\fL$ of the group manifold $\fL$.
A separation of variables in
(\ref{Complex}) is realized via the following factorization:
\begin{eqnarray}
\psi^k_{lm;\dot{l}\dot{m}}&=&\boldsymbol{f}^{l_0}_{lmk}(r)
\fM_{l_0}^{m}(\varphi,\epsilon,\theta,\tau,0,0),\nonumber\\
\psi^{\dot{k}}_{\dot{l}\dot{m};lm}&=&
\boldsymbol{f}^{\dot{l}_0}_{\dot{l}\dot{m}\dot{k}}(r^\ast)
\fM_{\dot{l}_0}^{\dot{m}}(\varphi,\epsilon,\theta,\tau,0,0),\label{F}
\end{eqnarray}
where $l_0\ge l$, $-l_0\le m$ and $\dot{l}_0\ge\dot{l}$,
$-\dot{l}_0\le\dot{m}$, $\fM_{l_0}^m(\varphi,\epsilon,\theta,\tau,0,0)$
($\fM_{\dot{l}}^{\dot{m}}(\varphi,\epsilon,\theta,\tau,0,0)$) are
associated hyperspherical functions defined on the surface of the 
two-dimensional complex sphere of the radius $r$, 
$\boldsymbol{f}^{l_0}_{lmk}(r)$ and
$\boldsymbol{f}^{\dot{l}_0}_{\dot{l}\dot{m}\dot{k}}(r^\ast)$ are radial
functions (for more details on two-dimensional complex sphere see
\cite{HS70,SH70,Hus71,Var031}). 
The associated hyperspherical function $\fM^m_l$ has a form
\[
\fM^m_l(\varphi,\epsilon,\theta,\tau,0,0)=e^{-m(\epsilon+i\varphi)}
Z^l_m(\theta,\tau),
\]
where the function $Z^l_m$ can be represented by a product of the two
hypergeometric functions:
\begin{multline}
Z^l_{mn}(\theta,\tau)=\cos^{2l}\frac{\theta}{2}\ch^{2l}\frac{\tau}{2}
\sum^l_{k=-l}i^{m-k}\tg^{m-k}\frac{\theta}{2}
\tnh^{-k}\frac{\tau}{2}\times\\[0.2cm]
\hypergeom{2}{1}{m-l+1,1-l-k}{m-k+1}{i^2\tg^2\dfrac{\theta}{2}}
\hypergeom{2}{1}{-l+1,1-l-k}{-k+1}{\tnh^2\dfrac{\tau}{2}}.\nonumber
\end{multline}

\section{The field $(1/2,0)\oplus(0,1/2)$}
Let us consider now an explicit construction of the relativistic
wavefunction $\psi(\balpha)$ on the Poincar\'{e} group for the first
nontrivial case described by the field $(1/2,0)\oplus(0,1/2)$ 
(electron-positron field or Dirac field). In this case, the first equation
from (\ref{FET}) coincides with the Dirac equation
\begin{equation}\label{Dirac}
i\gamma_nu\frac{\partial\psi(x)}{\partial x_\nu}-m\psi(x)=0,
\end{equation}
where $\gamma$-matrices are defined in the standard form, that is,
in the Weyl basis:
\[
\gamma_0=\begin{pmatrix}
\sigma_0 & 0\\
0 & -\sigma_0
\end{pmatrix},\;\;\gamma_1=\begin{pmatrix}
0 & \sigma_1\\
-\sigma_1 & 0
\end{pmatrix},\;\;\gamma_2=\begin{pmatrix}
0 & \sigma_2\\
-\sigma_2 & 0
\end{pmatrix},\;\;\gamma_3=\begin{pmatrix}
0 & \sigma_3\\
-\sigma_3 & 0
\end{pmatrix},
\]
where $\sigma_i$ are the Pauli matrices.

As is known, solutions of the equation (\ref{Dirac}) are found in the
plane-wave approximation, that is, in the form \cite{BD64,Ryd85}
\begin{eqnarray}
\psi^+(x)&=&u(\bp)e^{-ipx},\nonumber\\
\psi^-(x)&=&v(\bp)e^{ipx},\nonumber
\end{eqnarray}
where the solutions $\psi^+(x)$ and $\psi^-(x)$ correspond to positive
and negative energy, respectively, and the amplitudes $u(\bp)$ and
$v(\bp)$ have the following components:
\begin{gather}
u_1(\bp)=\left(\frac{E+m}{2m}\right)^{1/2}\begin{bmatrix}
1\\
0\\
\frac{p_z}{E+m}\\
\frac{p_+}{E+m}
\end{bmatrix},\quad
u_2(\bp)=\left(\frac{E+m}{2m}\right)^{1/2}\begin{bmatrix}
0\\
1\\
\frac{p_-}{E+m}\\
\frac{-p_z}{E+m}
\end{bmatrix},\nonumber\\
v_1(\bp)=\left(\frac{E+m}{2m}\right)^{1/2}\begin{bmatrix}
\frac{p_z}{E+m}\\
\frac{p_+}{E+m}\\
1\\
0
\end{bmatrix},\quad
v_2(\bp)=\left(\frac{E+m}{2m}\right)^{1/2}\begin{bmatrix}
\frac{p_-}{E+m}\\
\frac{-p_z}{E+m}\\
0\\
1
\end{bmatrix},\nonumber
\end{gather}
where $p_\pm=p_x\pm ip_y$.

Let us consider now solutions of the system
(\ref{Complex}) for the spin $l=1/2$, that is, when the field is defined by
a $P$-invariant direct sum $(1/2,0)\oplus(0,1/2)$. In this case the matrices
$\sL_i$ and $\sL^\ast_i$ have the form
\begin{eqnarray}
&&\sL_1=\frac{1}{2}c_{\frac{1}{2}\frac{1}{2}}\begin{pmatrix}
0 & 1\\
1 & 0
\end{pmatrix},\quad
\sL_2=\frac{1}{2}c_{\frac{1}{2}\frac{1}{2}}\begin{pmatrix}
0 & -i\\
i & 0
\end{pmatrix},\quad
\sL_3=\frac{1}{2}c_{\frac{1}{2}\frac{1}{2}}\begin{pmatrix}
1 & 0\\
0 & -1
\end{pmatrix},\nonumber\\
&&\sL^\ast_1=\frac{1}{2}\dot{c}_{\frac{1}{2}\frac{1}{2}}\begin{pmatrix}
0 & 1\\
1 & 0
\end{pmatrix},\quad
\sL^\ast_2=\frac{1}{2}\dot{c}_{\frac{1}{2}\frac{1}{2}}\begin{pmatrix}
0 & -i\\
i & 0
\end{pmatrix},\quad
\sL^\ast_3=\frac{1}{2}\dot{c}_{\frac{1}{2}\frac{1}{2}}\begin{pmatrix}
1 & 0\\
0 & -1
\end{pmatrix}.\nonumber
\end{eqnarray}
It is easy to see that these matrices coincide with the Pauli matrices
$\sigma_i$ when $c_{\frac{1}{2}\frac{1}{2}}=2$. The system (\ref{Complex})
at $l=1/2$ and 
$c_{\frac{1}{2},\frac{1}{2}}=\dot{c}_{\frac{1}{2},\frac{1}{2}}$ takes the form
\begin{eqnarray}
&&-\frac{1}{2}\frac{\partial\dot{\psi}_2}{\partial\widetilde{a}_1}+
\frac{i}{2}\frac{\partial\dot{\psi}_2}{\partial\widetilde{a}_2}-
\frac{1}{2}\frac{\partial\dot{\psi}_1}{\partial\widetilde{a}_3}-
\frac{i}{2}\frac{\partial\dot{\psi}_2}{\partial\widetilde{a}^\ast_1}-
\frac{1}{2}\frac{\partial\dot{\psi}_2}{\partial\widetilde{a}^\ast_2}-
\frac{i}{2}\frac{\partial\dot{\psi}_1}{\partial\widetilde{a}^\ast_3}-
\kappa^c\psi_1=0,\nonumber\\
&&-\frac{1}{2}\frac{\partial\dot{\psi}_1}{\partial\widetilde{a}_1}-
\frac{i}{2}\frac{\partial\dot{\psi}_1}{\partial\widetilde{a}_2}+
\frac{1}{2}\frac{\partial\dot{\psi}_2}{\partial\widetilde{a}_3}-
\frac{i}{2}\frac{\partial\dot{\psi}_1}{\partial\widetilde{a}^\ast_1}+
\frac{1}{2}\frac{\partial\dot{\psi}_1}{\partial\widetilde{a}^\ast_2}+
\frac{i}{2}\frac{\partial\dot{\psi}_2}{\partial\widetilde{a}^\ast_3}-
\kappa^c\psi_2=0,\nonumber\\
&&\phantom{-}\frac{1}{2}\frac{\partial\psi_2}{\partial a_1}-
\frac{i}{2}\frac{\partial\psi_2}{\partial a_2}+
\frac{1}{2}\frac{\partial\psi_1}{\partial a_3}-
\frac{i}{2}\frac{\partial\psi_2}{\partial a^\ast_1}-
\frac{1}{2}\frac{\partial\psi_2}{\partial a^\ast_2}-
\frac{i}{2}\frac{\partial\psi_1}{\partial a^\ast_3}-
\dot{\kappa}^c\dot{\psi}_1=0,\nonumber\\
&&\phantom{-}\frac{1}{2}\frac{\partial\psi_1}{\partial a_1}+
\frac{i}{2}\frac{\partial\psi_1}{\partial a_2}-
\frac{1}{2}\frac{\partial\psi_2}{\partial a_3}-
\frac{i}{2}\frac{\partial\psi_1}{\partial a^\ast_1}+
\frac{1}{2}\frac{\partial\psi_1}{\partial a^\ast_2}+
\frac{i}{2}\frac{\partial\psi_2}{\partial a^\ast_3}-
\dot{\kappa}^c\dot{\psi}_2=0,\label{Dirac2}
\end{eqnarray}
This system is defined on the tangent bundle $T\fL$ of the group
manifold $\fL$. Let us find solutions of the system (\ref{Dirac2}) in terms
of the functions on the Lorentz group:
\begin{eqnarray}
\psi_1(\fg)&=&
\boldsymbol{f}^l_{\frac{1}{2},\frac{1}{2}}(r)
\fM_l^{\frac{1}{2}}(\varphi,\epsilon,\theta,\tau,0,0),\nonumber\\
\psi_2(\fg)&=&
\boldsymbol{f}^l_{\frac{1}{2},-\frac{1}{2}}(r)
\fM_l^{-\frac{1}{2}}(\varphi,\epsilon,\theta,\tau,0,0),\nonumber\\
\dot{\psi}_1(\fg)&=&
\boldsymbol{f}^{\dot{l}}_{\frac{1}{2},\frac{1}{2}}(r^\ast)
\fM_{\dot{l}}^{\frac{1}{2}}(\varphi,\epsilon,\theta,\tau,0,0),\nonumber\\
\dot{\psi}_2(\fg)&=&
\boldsymbol{f}^{\dot{l}}_{\frac{1}{2},-\frac{1}{2}}(r^\ast)
\fM_{\dot{l}}^{-\frac{1}{2}}(\varphi,\epsilon,\theta,\tau,0,0),\nonumber
\end{eqnarray}
Substituting these functions into (\ref{Dirac2}) and separating the variables
with the aid of recurrence relations between hyperspherical functions,
we come to the following system of ordinary differential equations:
\begin{eqnarray}
&&-\frac{1}{2}\frac{d\boldsymbol{f}^{\dot{l}}_{\frac{1}{2},\frac{1}{2}}(r^\ast)}
{d r^\ast}+\frac{1}{4r^\ast}\boldsymbol{f}^{\dot{l}}_{\frac{1}{2},\frac{1}{2}}
(r^\ast)+\frac{\dot{l}+\frac{1}{2}}{2r^\ast}
\boldsymbol{f}^{\dot{l}}_{\frac{1}{2},-\frac{1}{2}}(r^\ast)-
\kappa^c\boldsymbol{f}^l_{\frac{1}{2},\frac{1}{2}}(r)=0,\nonumber\\
&&\phantom{-}\frac{1}{2}\frac{d\boldsymbol{f}^{\dot{l}}_{\frac{1}{2},-\frac{1}{2}}(r^\ast)}
{d r^\ast}-\frac{1}{4r^\ast}\boldsymbol{f}^{\dot{l}}_{\frac{1}{2},-\frac{1}{2}}
(r^\ast)-\frac{\dot{l}+\frac{1}{2}}{2r^\ast}
\boldsymbol{f}^{\dot{l}}_{\frac{1}{2},\frac{1}{2}}(r^\ast)-
\kappa^c\boldsymbol{f}^l_{\frac{1}{2},-\frac{1}{2}}(r)=0,\nonumber\\
&&\phantom{-}\frac{1}{2}\frac{d\boldsymbol{f}^l_{\frac{1}{2},\frac{1}{2}}(r)}{dr}-
\frac{1}{4r}\boldsymbol{f}^l_{\frac{1}{2},\frac{1}{2}}(r)-
\frac{l+\frac{1}{2}}{2r}\boldsymbol{f}^l_{\frac{1}{2},-\frac{1}{2}}-
\dot{\kappa}^c\boldsymbol{f}^{\dot{l}}_{\frac{1}{2},\frac{1}{2}}(r^\ast)=0,
\nonumber\\
&&-\frac{1}{2}\frac{d\boldsymbol{f}^l_{\frac{1}{2},-\frac{1}{2}}(r)}{dr}+
\frac{1}{4r}\boldsymbol{f}^l_{\frac{1}{2},-\frac{1}{2}}(r)+
\frac{l+\frac{1}{2}}{2r}\boldsymbol{f}^l_{\frac{1}{2},\frac{1}{2}}-
\dot{\kappa}^c\boldsymbol{f}^{\dot{l}}_{\frac{1}{2},-\frac{1}{2}}(r^\ast)=0,
\nonumber
\end{eqnarray}
For the brevity of exposition we suppose 
$\boldsymbol{f}_1=\boldsymbol{f}^l_{\frac{1}{2},\frac{1}{2}}(r)$,
$\boldsymbol{f}_2=\boldsymbol{f}^l_{\frac{1}{2},-\frac{1}{2}}(r)$,
$\boldsymbol{f}_3=\boldsymbol{f}^{\dot{l}}_{\frac{1}{2},\frac{1}{2}}(r^\ast)$,
$\boldsymbol{f}_4=\boldsymbol{f}^{\dot{l}}_{\frac{1}{2},-\frac{1}{2}}(r^\ast)$.
Then
\begin{eqnarray}
&&-2\frac{d\boldsymbol{f}_3}{dr^\ast}+\frac{1}{r^\ast}\boldsymbol{f}_3+
\frac{2\left(\dot{l}+\frac{1}{2}\right)}{r^\ast}\boldsymbol{f}_4-
4\kappa^c\boldsymbol{f}_1=0,\nonumber\\
&&\phantom{-}2\frac{d\boldsymbol{f}_4}{dr^\ast}-\frac{1}{r^\ast}\boldsymbol{f}_4-
\frac{2\left(\dot{l}+\frac{1}{2}\right)}{r^\ast}\boldsymbol{f}_3-
4\kappa^c\boldsymbol{f}_2=0,\nonumber\\
&&\phantom{-}2\frac{d\boldsymbol{f}_1}{dr}-\frac{1}{r}\boldsymbol{f}_1-
\frac{2\left(l+\frac{1}{2}\right)}{r}\boldsymbol{f}_2-
4\dot{\kappa}^c\boldsymbol{f}_3=0,\nonumber\\
&&-2\frac{d\boldsymbol{f}_2}{dr}+\frac{1}{r}\boldsymbol{f}_2+
\frac{2\left(l+\frac{1}{2}\right)}{r}\boldsymbol{f}_1-
4\dot{\kappa}^c\boldsymbol{f}_4=0.\nonumber
\end{eqnarray}
Let us assume that $\boldsymbol{f}_3=\mp\boldsymbol{f}_4$ and
$\boldsymbol{f}_2=\pm\boldsymbol{f}_1$; then the first equation coincides
with the second, and the third equations coincides with the fourth.
Therefore,
\begin{eqnarray}
&&\frac{d\boldsymbol{f}_4}{dr^\ast}+\frac{\dot{l}}{r^\ast}\boldsymbol{f}_4-
2\kappa^c\boldsymbol{f}_1=0,\nonumber\\
&&\frac{d\boldsymbol{f}_1}{dr}-\frac{l+1}{r}\boldsymbol{f}_1+
2\dot{\kappa}^c\boldsymbol{f}_4=0.\nonumber
\end{eqnarray}
Let us consider a real part $\re r$ of the radius of a complex sphere.
It is obvious that $\re r=\re r^\ast$. Writing $\sz=\re r=\re r^\ast$ and
excluding the function $\boldsymbol{f}_4$ at $l=\dot{l}$, 
we come to the following differential equation:
\begin{equation}\label{Bess}
\sz^2\frac{d^2\boldsymbol{f}_1}{d\sz^2}-\sz\frac{d\boldsymbol{f}_1}{d\sz}-
(l^2-1-4\kappa^c\dot{\kappa}^c\sz^2)\boldsymbol{f}_1=0.
\end{equation}
The latter equation is solvable in the Bessel functions of half-integer
order:
\[
\boldsymbol{f}_1(\sz)=C_1\sqrt{\kappa^c\dot{\kappa}^c}\sz 
J_l\left(\sqrt{\kappa^c\dot{\kappa}^c}\sz\right)+
C_2\sqrt{\kappa^c\dot{\kappa}^c}\sz
J_{-l}\left(\sqrt{\kappa^c\dot{\kappa}^c}\sz\right).
\]
Further, using recurrence relations between Bessel functions, we find
\begin{multline}
\boldsymbol{f}_4(\sz)=\frac{1}{2\kappa^c}
\left(\frac{l+1}{\sz}\boldsymbol{f}_1(\sz)-
\frac{d\boldsymbol{f}_1}{d\sz}\right)=\\
=\frac{C_1}{2}\sqrt{\frac{\dot{\kappa}^c}{\kappa^c}}\sz
J_{l+1}\left(\sqrt{\kappa^c\dot{\kappa}^c}\sz\right)-
\frac{C_2}{2}\sqrt{\frac{\dot{\kappa}^c}{\kappa^c}}\sz
J_{-l-1}\left(\sqrt{\kappa^c\dot{\kappa}^c}\sz\right).\nonumber
\end{multline}
Therefore,
\[
\boldsymbol{f}^l_{\frac{1}{2},\frac{1}{2}}(\re r)=
C_1\sqrt{\kappa^c\dot{\kappa}^c}\re r 
J_l\left(\sqrt{\kappa^c\dot{\kappa}^c}\re r\right)+
C_2\sqrt{\kappa^c\dot{\kappa}^c}\re r 
J_{-l}\left(\sqrt{\kappa^c\dot{\kappa}^c}\re r\right),
\]
\[
\boldsymbol{f}^{\dot{l}}_{\frac{1}{2},-\frac{1}{2}}(\re r^\ast)=
\frac{C_1}{2}\sqrt{\frac{\dot{\kappa}^c}{\kappa^c}}\re r^\ast
J_{l+1}\left(\sqrt{\kappa^c\dot{\kappa}^c}\re r^\ast\right)
-\frac{C_2}{2}\sqrt{\frac{\dot{\kappa}^c}{\kappa^c}}\re r^\ast
J_{-l-1}\left(\sqrt{\kappa^c\dot{\kappa}^c}\re r^\ast\right).
\]
%where $\nu=-(l-1)$, $\dot{\nu}=-(\dot{l}-1)$, 
%$l=\frac{2s+1}{2}$, $\dot{l}=\frac{2\dot{s}+1}{2}$,
%$s,\dot{s}=0,1,2,\ldots$ and
%\begin{multline}
%J_{\frac{2s+1}{2}}(z)=\sqrt{\frac{2}{\pi z}}\left[
%\sin\left(z-\frac{s\pi}{2}\right)\sum^{\frac{s}{2}}_{k=0}
%\frac{(-1)^k(s+2k)!}{(2k)!(s-2k)!(2z)^{2k}}+\right.\\
%\left.
%+\cos\left(z-\frac{s\pi}{2}\right)\sum^{\left[\frac{s-1}{2}\right]}_{k=0}
%\frac{(-1)^k(s+2k+1)!}{(2k+1)!(s-2k-1)!(2z)^{2k+1}}\right].\label{Bessel}
%\end{multline}
%is the Bessel function of half-integer order. 
In this way, solutions of the system (\ref{Dirac2})
are defined by the following functions:
\begin{eqnarray}
\psi_1(r,\varphi^c,\theta^c)&=&\boldsymbol{f}^l_{\frac{1}{2},\frac{1}{2}}
(\re r)\fM_l^{\frac{1}{2}}(\varphi,\epsilon,\theta,\tau,0,0),\nonumber\\
\psi_2(r,\varphi^c,\theta^c)&=&\pm\boldsymbol{f}^l_{\frac{1}{2},\frac{1}{2}}
(\re r)\fM_l^{-\frac{1}{2}}(\varphi,\epsilon,\theta,\tau,0,0),\nonumber\\
\dot{\psi}_1(r^\ast,\dot{\varphi}^c,\dot{\theta}^c)&=&
\mp\boldsymbol{f}^{\dot{l}}_{\frac{1}{2},-\frac{1}{2}}
(\re r^\ast)\fM_{\dot{l}}^{\frac{1}{2}}
(\varphi,\epsilon,\theta,\tau,0,0),\nonumber\\
\dot{\psi}_2(r^\ast,\dot{\varphi}^c,\dot{\theta}^c)&=&
\boldsymbol{f}^{\dot{l}}_{\frac{1}{2},-\frac{1}{2}}
(\re r^\ast)\fM_l^{-\frac{1}{2}}(\varphi,\epsilon,\theta,\tau,0,0),
\nonumber
\end{eqnarray}
where
\begin{eqnarray}
&&l=\frac{1}{2},\;\frac{3}{2},\;\frac{5}{2},\ldots;\nonumber\\
&&\dot{l}=\frac{1}{2},\;\frac{3}{2},\;\frac{5}{2},\ldots;\nonumber
\end{eqnarray}
\[
\fM_l^{\pm\frac{1}{2}}(\varphi,\epsilon,\theta,\tau,0,0)=
e^{\mp\frac{1}{2}(\epsilon+i\varphi)}Z_l^{\pm\frac{1}{2}}(\theta,\tau),
\]
\begin{multline}
Z_l^{\pm\frac{1}{2}}(\theta,\tau)=\cos^{2l}\frac{\theta}{2}
\ch^{2l}\frac{\tau}{2}\sum^l_{k=-l}i^{\pm\frac{1}{2}-k}
\tg^{\pm\frac{1}{2}-k}\frac{\theta}{2}\tnh^{-k}\frac{\tau}{2}\times\\
\hypergeom{2}{1}{\pm\frac{1}{2}-l+1,1-l-k}{\pm\frac{1}{2}-k+1}
{i^2\tg^2\frac{\theta}{2}}
\hypergeom{2}{1}{-l+1,1-l-k}{-k+1}{\tnh^2\frac{\tau}{2}},\nonumber
\end{multline}
\[
\fM_{\dot{l}}^{\pm\frac{1}{2}}(\varphi,\epsilon,\theta,\tau,0,0)=
e^{\mp\frac{1}{2}(\epsilon-i\varphi)}
Z_{\dot{l}}^{\pm\frac{1}{2}}(\theta,\tau),
\]
\begin{multline}
Z_{\dot{l}}^{\pm\frac{1}{2}}(\theta,\tau)=
\cos^{2\dot{l}}\frac{\theta}{2}
\ch^{2\dot{l}}\frac{\tau}{2}
\sum^{\dot{l}}_{\dot{k}=-\dot{l}}i^{\pm\frac{1}{2}-\dot{k}}
\tg^{\pm\frac{1}{2}-\dot{k}}\frac{\theta}{2}
\tnh^{-\dot{k}}\frac{\tau}{2}\times\\
\hypergeom{2}{1}{\pm\frac{1}{2}-\dot{l}+1,1-\dot{l}-\dot{k}}
{\pm\frac{1}{2}-\dot{k}+1}
{i^2\tg^2\frac{\theta}{2}}
\hypergeom{2}{1}{-\dot{l}+1,1-\dot{l}-\dot{k}}
{-\dot{k}+1}{\tnh^2\frac{\tau}{2}}.\nonumber
\end{multline}

Therefore, in accordance with the factorization (\ref{WF}), an explicit
form of the relativistic wavefunction $\psi(\balpha)=\psi(x)\psi(\fg)$
on the Poincar\'{e} group in the case of 
$(1/2,0)\oplus(0,1/2)$-representation is given
by the following expressions:
\begin{eqnarray}
&&\psi_1(\balpha)=\psi^+_1(x)\psi_1(\fg)=
u_1(\bp)e^{-ipx}\boldsymbol{f}^l_{\frac{1}{2},\frac{1}{2}}
(\re r)\fM_l^{\frac{1}{2}}(\varphi,\epsilon,\theta,\tau,0,0),\nonumber\\
&&\psi_2(\balpha)=\psi^+_2(x)\psi_2(\fg)=
\pm u_2(\bp)e^{-ipx}\boldsymbol{f}^l_{\frac{1}{2},\frac{1}{2}}
(\re r)\fM_l^{-\frac{1}{2}}(\varphi,\epsilon,\theta,\tau,0,0),\nonumber\\
&&\dot{\psi}_1(\balpha)=\psi^-_1(x)\dot{\psi}_1(\fg)=
\mp v_1(\bp)e^{ipx}\boldsymbol{f}^{\dot{l}}_{\frac{1}{2},-\frac{1}{2}}
(\re r^\ast)\fM_{\dot{l}}^{\frac{1}{2}}
(\varphi,\epsilon,\theta,\tau,0,0),\nonumber\\
&&\dot{\psi}_2(\balpha)=\psi^-_2(x)\dot{\psi}_2(\fg)=
v_2(\bp)e^{ipx}\boldsymbol{f}^{\dot{l}}_{\frac{1}{2},-\frac{1}{2}}
(\re r^\ast)\fM_l^{-\frac{1}{2}}(\varphi,\epsilon,\theta,\tau,0,0),
\label{WF2}
\end{eqnarray}
The quantities (\ref{WF2}) form {\it a bispinor on the Poincar\'{e} group},
$\psi(\balpha)=(\psi_1(\balpha),\psi_2(\balpha),\dot{\psi}_1(\balpha),
\dot{\psi}_2(\balpha))^T$.

It is obvious that, solving the equations for $\psi(x)$ and
$\psi(\fg)$ separately, 
we can find in like manner all the parametrized forms of the
functions (\ref{SF2}) for any spin. The important case of the field
$(1,0)\oplus(0,1)$ will be considered in a separate paper.

In conclusion, it should be noted the following circumstance.
As is known, in the standard QFT, solutions of relativistic wave equations
are found in the plane-wave approximation (it is hardly too much to say that
these solutions are strongly degenerate) and field operators are defined
in the form of Fourier expansions (or Fourier integrals) in such solutions.
In passing to the functions on the Poincar\'{e} group and their
parametrized forms, the Fourier expansions are replaced by more general
transformations; that is, we come to an expansion of $\psi(\balpha)$ on the
group $\cP$ or harmonic analysis of the functions on the groups
(see, for example, \cite{Ruh70,Hai69,Hai71,Str73,She75}). In this way, usual
Fourier analysis of the standard QFT is replaced by harmonic analysis in the
case of QFTPG. The field operators on the functions (\ref{WF}) and their
parametrizations of the form (\ref{WF2}) will be studied in the future
work in terms of harmonic analysis on the Poincar\'{e} group.

\end{document}